\documentclass[10pt,showpacs,showkeys,preprintnumbers]{article}
\usepackage{amsmath,amssymb,amsfonts}
\usepackage{units}
\usepackage{bbold,euler}
\usepackage{graphicx}
\usepackage{dcolumn}
\usepackage{bm,bbm,mathtools,esvect}
\usepackage{slashed}
\usepackage{esvect}
\usepackage{fullpage}
\usepackage{comment}
\usepackage[utf8]{inputenc}
\title{\bf ATTENUATED GRAVITATIONAL RADIATION }
\author{\\ SERGIO GIARDINO\footnote{\tt sergio.giardino@ufrgs.br}\\
\\
\small \it Departamento de Matem\'atica Pura e Aplicada \\
\small \it Universidade Federal do Rio Grande do Sul (UFRGS)\\
\small \it Caixa Postal 15080, 91501-970  Porto Alegre RS \\
\small \it Brazil}

\begin{document}
\date{}
\maketitle

\begin{abstract}
\noindent 
The hypothesis of an alternative way of obtaining gravitational waves is the physical motivation of this article. 
Using the linear field approximation and a symmetry transformation of  the affine connection, new field equations and new gauge conditions have been obtained. Solutions to these field equations have been considered in the empty space and in the wave zone, and in both of them the oscillation amplitudes of their solutions are attenuated exponentially. We expect that these solutions can be useful for building more sophisticated gravitational wave models, and also as an impulse for researching further symmetry transformations of general relativity.
\end{abstract}


\section{\;\sc Introduction\label{I}}

Gravitational waves are generated after sudden changes concerning the distribution of matter and energy. Violent events like the collision of black holes and 
explosion of supernovae are examples of phenomena that can generate detectable gravitational waves. These events perturb the geometry of the space time in a manner that a generates dynamical wave whose motion is described using the weak field description of general relativity. Within this approximation, the metric tensor $\,g_{\mu\nu}\,$ is proposed to be
\begin{equation}\label{i00}
g_{\mu\nu}=\eta_{\mu\nu}+ h_{\mu\nu}, \qquad\mbox{where}\qquad |h_{\mu\nu}|\ll 1,
\end{equation}
and $\,\eta_{\mu\nu}\,$ is the metric tensor of the plane Minkowski space with signature $\,(+\,-\,-\,-).$
It can be proven \cite{Weinberg:1972kfs,Ohanian:1995uu,dInverno:1992gxs} that the tensor field $\,h_{\mu\nu}\,$ that describes the metric effect due to the change in the mass and energy distribution satisfies the wave equation, and hence rendering the gravitational wave phenomenon. In this article, we pose the question about whether (\ref{i00}) alone is the unique way for perturbing the space-time, and
we propose a method to perturb the space-time that comprises the local perturbation of the metric tensor (\ref{i00}) and a global transformation of the affine connection. In other words, we highlight that the motivation of this study is mainly theoretical, and is not intended to explain an observed phenomenon. Instead, our aim is to transform the theory in order to obtain a more flexible framework that will be able do describe a wider range of phenomena in the scope of gravitational radiation.

Before get into the details, we remember that gravitational radiation is one of the most active fields in contemporary research of gravity, particularly after the experimental verification of this phenomenon \cite{LIGOScientific:2018mvr,Maggiore:2018sht,Cervantes-Cota:2016zjc}, but the subject is much wider, as can be seen from books and reviews on the subject \cite{Corda:2009re,Maggiore:1900zz,Nielsen:2019caa}. The range of studies in gravitational waves (GW) is so large and so diverse that it became virtually independent of General Relativity. 

 As we shall see, in this paper we propose a geometric change in general relativity that will result in attenuated gravitational waves. The hypothesis of  attenuation and amplification of GW is a very recent subject, and we mention the damped GW in dark matter cosmologies as a first reference \cite{Bian:2021ini}. Otherwise, the space-time decay \cite{Bieri:2020pee,Bieri:2020zki} presents various novel structures in GW phenomenology, including memory and the decaying of neutrino radiation. In quantum gravity, the vanishing of graviton amplitudes \cite{Gamboa:2020yqh} also indicates the importance of attenuated GW, and finally the gravitomagnetic effects of GW also generate ressonance interactions \cite{Ruggiero:2021uag}. In summary, the idea of varying amplitudes of GW is certainly not introduced here, although is an emergent matter of investigation. The novelty  is the proposition of the geometric model that describes it, and this is the main importance contained in this paper.

Therefore, we investigate a different theoretical method of obtaining the gravitational radiation, and this novel method generate solutions that are different from the usual GW solutions. The method consist of a global and rigid transformation of the space-time obtained by transforming the affine connection. The transformation we will propose demands the symmetric tensor $\,\tau^{\mu\nu},\,$ whose components are
\begin{equation}\label{i01}
\tau^{\;\;\;\mu}_\nu=\left\{
\begin{array}{ll}
\;\;\;\, 1, &\quad\mbox{if}\quad \mu=\nu=0\\
&\\
-1, &\quad\mbox{if}\quad\mu=\nu=i\\
&\\
\;\;\;\,0, &\quad\mbox{if}\quad\mu\neq\nu.
\end{array}
\right.
\end{equation}
where $\,\mu,\,\nu\,$ are space-time indexes and $\,i,\,j\,$ are space indexes. The above tensor also satisfies
\begin{equation}\label{i02}
g_{\mu\nu}\,=\,g_{\kappa\lambda}\,\tau^{\;\;\;\kappa}_\mu\,\tau^{\;\;\;\lambda}_\nu\qquad\mbox{in the case where}\qquad g_{0i}=0.
\end{equation}
We notice that $\,\tau^{\mu\nu}\,$ was proposed for a plane space in a study of gravito-electromagnetism \cite{Giardino:2018ffd,Giardino:2021gwq}, and (\ref{i01}-\ref{i02}) generalize that previous definition for a curved space. Let us deform then the affine connection as
\begin{equation}\label{i03}
\Gamma^\lambda_{\mu\nu}\quad\to\quad \tau^\lambda_{\;\;\;\kappa}\Gamma^\kappa_{\mu\nu}\,=\,
\frac{1}{2}\tau^{\lambda\kappa}\Bigg(\partial_\mu g_{\kappa\nu}+\partial_\nu g_{\kappa\mu}-\partial_\kappa g_{\mu\nu}\Bigg).
\end{equation}
This global transformation changes the parallel transportation of a vector along the geodesic line. One can ask the physical motivation of (\ref{i03}). From the theoretical point of view, it is a mathematical generalization of (\ref{i00}), and it may open the way to a further and more profound generalization. From the physical point of view, there is any physical evidence on how the perturbation process takes place. Although the proposal (\ref{i00}) is  simple and reasonable, it is impossible to assure that this is the unique possible way to perturb the space-time, and  it seems improbable to expect that every different physical process will deform the space-time in a way that could be described in this single form. We therefore investigate an alternative manner for perturbing the space-time, and hope that further possibilities will rise up from the future directions of research.

We could simply derive the consequences of this transformation using the linear approximation. However, let us first analyze the meaning and the geometrical consequences of (\ref{i03}). First of all, we observe that the effect of the transformation is flip the signal of the affine connection as
\begin{equation}\label{i04}
\Gamma^\lambda_{\mu\nu}\quad\to\quad \tau^\lambda_{\;\;\;\kappa}\Gamma^\kappa_{\mu\nu}\qquad\Rightarrow\qquad
\Big(\Gamma^0_{\mu\nu},\,\Gamma^i_{\mu\nu}\Big)\quad\to\quad \Big(\Gamma^0_{\mu\nu},\,-\,\Gamma^i_{\mu\nu}\Big)
\end{equation}
If we remind the definition of the covariant derivative,
\begin{equation}\label{i05}
\nabla_\mu v^\nu=\partial_\mu v^\nu + \Gamma^\nu_{\mu\lambda}v^\lambda,\qquad \qquad
\nabla_\mu v_\nu=\partial_\mu v_\nu - \Gamma^\lambda_{\mu\nu}v_\lambda 
\end{equation}
we observe that the transformation (\ref{i04}) make the spatial components $\,v^i\,$ transform in the opposite direction of the usual
parallel transport. Additionally, we observe that
\begin{equation}\label{i0005}
 \nabla_\lambda g_{\mu\nu}=\left(\Gamma^\rho_{\mu\lambda}-\tau^\rho_{\;\sigma}\Gamma^\sigma_{\mu\lambda}\right)g_{\nu\rho}+
 \Big(\Gamma^\rho_{\nu\lambda}-\tau^\rho_{\;\sigma}\Gamma^\sigma_{\nu\lambda}\Big)g_{\mu\rho},
\end{equation}
and the quantity in the right hand side of (\ref{i0005}) is not identically zero, differently to what is observed in Riemannian geometry. Therefore, transformation (\ref{i03})  indicates the breakdown of the Lorentz symmetry, and the emergence of an affine gravitation space-time \cite{Hehl:1994ue}. However, this broken symmetry may be avoided by imposing (\ref{i0005}) to be identically zero as a constraint, and this is certainly an interesting direction for future research. Moreover, (\ref{i0005}) means that the covariant derivative, and the operations or rising and lowering indices using the metric tensor do not commute, 
and the affine connection has to be corrected \cite{Hehl:1994ue,Jarv:2018bgs}  as
\begin{equation}\label{AC}
 \Gamma^\mu_{\nu\kappa}\;\to\; \Gamma^\mu_{\nu\kappa}+K^\mu_{\nu\kappa}+L^\mu_{\nu\kappa},
\end{equation}
where $K^\mu_{\nu\kappa}$ and $L^\mu_{\nu\kappa}$ are respectively the contortion and disformation terms. Moreover, the Riemann tensor accordingly transforms such as
\begin{equation}\label{Rie}
 R^\sigma_{\;\;\rho\mu\nu}\;\to\; R^\sigma_{\;\;\rho\mu\nu}+\nabla_\mu M^\sigma_{\nu\rho}-\nabla_\nu M^\sigma_{\mu\rho} + M^\lambda_{\mu\rho} M^\sigma_{\nu\lambda}-  M^\lambda_{\nu\rho} M^\sigma_{\mu\lambda},
\end{equation}
where $M^\sigma_{\nu\kappa}=K^\mu_{\nu\kappa}+L^\mu_{\nu\kappa}$. This complete transformation will not be considered in the sequel, and the analysis will be focused in the leading term only, and we will set $M^\lambda_{\nu\rho}=0$.  However, the more general situation, where $M^\lambda_{\nu\rho}\neq 0$,  must be seriously considered in future research, and the results of this paper can be considered the first approximation to this complete theory. 
Therefore, the transformed Riemann tensor is so that
\begin{equation}\label{rt01}
\mathcal{R}^\lambda_{\;\;\;\mu\nu\kappa}\;=\;\tau^\lambda_{\;\;\lambda'}\Bigg(\,\partial_\nu\Gamma^{\lambda'}_{\mu\kappa}\,-\,\partial_\kappa\Gamma^{\lambda'}_{\mu\nu}\,\Bigg)\,+\,\tau^\lambda_{\;\;\lambda'}\tau^\eta_{\;\;\eta'}\Bigg(\,\Gamma^{\eta'}_{\mu\kappa}\Gamma^{\lambda'}_{\nu\eta}\,-\,\Gamma^{\eta'}_{\mu\nu}\Gamma^{\lambda'}_{\kappa\eta}\,\Bigg).
\end{equation}
Consequently,
\begin{equation}\label{i005}
\mathcal R_{\lambda\mu\nu\kappa}\;=\;\tau^\rho_{\;\;\lambda}\Bigg(\,\,\partial^2_{\kappa\rho}g_{\mu\nu}\,-\,\partial^2_{\kappa\mu}g_{\rho\nu}\,+\,\partial^2_{\mu\nu}g_{\rho\kappa}\,-\,\partial^2_{\rho\nu}g_{\mu\kappa}\Bigg)\,+
\,g_{\eta\sigma}\Bigg(\Gamma^\eta_{\kappa\lambda}\Gamma^\sigma_{\mu\nu}\,-\,\Gamma^\eta_{\nu\lambda}\Gamma^\sigma_{\mu\kappa}\Bigg)
\end{equation}
where $\;\tau^\kappa_{\;\;\mu}\tau_{\kappa\nu}=g_{\mu\nu}\;$ has been used. The properties of the transformed tensor are
\begin{equation}\label{i06}
\mathcal R_{\mu\mu\nu\kappa}\;=0,\qquad\qquad\mathcal R_{\lambda\mu\nu\kappa}\;=\;-\mathcal R_{\lambda\mu\kappa\nu}\qquad\qquad\mbox{and}\qquad\qquad \mathcal R_{\lambda\mu\nu\kappa}\;+\;\mathcal R_{\lambda\nu\kappa\mu}\;+\;\mathcal R_{\lambda\kappa\mu\nu}\;=\;0.
\end{equation}
We notice that the transformed Riemann is not enabled with every property of the Riemann tensor and
\begin{equation}\label{i07}
R_{\lambda\mu\nu\kappa}\,=\,R_{\nu\kappa\lambda\mu}\,=\,-\,R_{\kappa\nu\lambda\mu},
\end{equation}
are  not properties of  $\,\mathcal R_{\nu\kappa\lambda\mu}.\,$ The Bianchi identities
are also still valid, and the transformed Ricci tensor is still symmetric, so that 
\begin{equation}\label{i08}
\mathcal R_{\mu\nu}\,=\,\mathcal R_{\nu\mu}.
\end{equation}
Another question that can be formulated is what metric tensor would generate the transformed affine connection. As a simple example, let us consider a two-dimensional case, where the transformed affine connection (\ref{i04})  can be obtained from a transformed metric, so that
\begin{equation}\label{i09}
ds^2\,=\,g_{00} dt^2\,-\,g_{11}	dx^2\qquad \to \qquad ds^2\,=\,g_{00}\,dt^2\,-\,\frac{1}{g_{11}}dx^2,
\end{equation}
where $\,g_{00}=g_{00}(t)\,$ and $\,g_{11}=g_{11}(x)\,$ are real functions. However, if both of the components of the metric tensor are functions of the same variable, there is not a transformed metric tensor that could generate the transformed affine connections.  As a further feature of the two dimensional case, we remember that the Riemann tensor has a single component in this case $\,R_{0101}\,$ \cite{Weinberg:1972kfs}. Conversely, the transformed Riemann tensor in two-dimensions satisfies:
\begin{equation}\label{i10}
\mathcal R_{0101}\,\neq\,\mathcal R_{1001},\qquad \mbox{in opposition to}\qquad R_{0101}=- R_{1001}.
\end{equation}
Thus, the transformed Riemann tensor comprises two components, while the Riemann tensor comprises only one component in two dimensions. This evidences the disruptive effect of the transformation  (\ref{i04}) on the geometry of the space-time, which is distorted in a global and non-geometric way and whose consequence is the geometric oscillation that will be discussed in the next section. We stress that the transformation does not depict a permanent change in space-time, but a restrictive phenomenon that lasts only enough to generate the space-time perturbation. 

 As a last comment, we clarify that this article does not constitute a chapter of a new theory of gravity, which would either replace general relativity or contain it as a particular case within a certain limit. Instead of this, this paper contains an example of how a consistent modification of general relativity can extend their explanatory capacities, and therefore enabling it to fit a wider set of data.

\section{\;\sc Linearized field equations\label{OG}}
In this section, we derive the linearized field equations following the usual approach of \cite{Weinberg:1972kfs} in the geometrical conventions of Riemann geometry of \cite{dInverno:1992gxs}.  The physical consistency of the theory presented in this section has already been ascertained in \cite{Giardino:2018ffd}, where conservation laws and the full expression of the energy momentum tensor have been examined.
In particular, the positive definiteness of the energy momentum tensor of gravitational waves is effectively guaranteed. Therefore, the results of this section have a physical content that clearly overtakes their simple mathematical expression. We will observe that the net effect of the transformation (\ref{i04}) comprises a novel field equation to the geometrical wave and a more strict gauge condition. We start remembering the usual Einstein  gravitational equations
\begin{equation}\label{og01}
G_{\mu\nu}\,=\,\kappa\,T_{\mu\nu},\qquad\mbox{where}\qquad
 G_{\mu\nu}\,=\,R_{\mu\nu}-\frac{1}{2}g_{\mu\nu}R \qquad\mbox{and}\qquad\kappa=\frac{8\pi G}{c^4},
\end{equation}
that could also be written as  	
\begin{equation}\label{og02}
R_{\mu\nu}\,=\,\kappa S_{\mu\nu}\qquad\mbox{where}\qquad S_{\mu\nu}\,=\,T_{\mu\nu}-\frac{1}{2}g_{\mu\nu}T\qquad\mbox{and}\qquad T\,=\,g^{\mu\nu}T_{\mu\nu}.
\end{equation}
Using the linear approximation (\ref{i00}) and the affine connection (\ref{i03}) we obtain
\begin{equation}\label{og03}
\Gamma_{\mu\nu}^\lambda=\frac{1}{2}\tau^{\lambda\kappa}\Big(\partial_\mu h_{\kappa\nu}+\partial_\nu h_{\kappa\mu}-\partial_\kappa h_{\mu\nu}\Big)\,+\,\mathcal{O}\left(h_{\alpha\beta}^2\right).
\end{equation}
Despising the $\mathcal O\left(h_{\alpha\beta}^2\right)$ terms, the Ricci tensor is
\begin{equation}\label{og04}
\mathcal R_{\mu\nu}\,=\,\frac{1}{2}\,\tau^{\kappa\lambda}\,\Big[\,2
\,\partial_\kappa\partial_{(\mu} h_{\nu)\lambda}-\partial_\mu\partial_\nu h_{\kappa\lambda}-\partial_\kappa\partial_\lambda h_{\mu\nu}
\Big]\qquad\mbox{where}\qquad u_{(\mu}v_{\nu)}=\frac{1}{2}\Big(u_\mu v_\nu+u_\nu v_\mu\Big)
\end{equation}
and the Ricci scalar is
\begin{equation}\label{og05}
\mathcal R\,=\,\frac{1}{2}\,\tau^{\kappa\lambda}\,\Big(\,
2\,\partial_\lambda\partial_\alpha h^\alpha_{\;\;\kappa}-\partial_\kappa\partial_\lambda h-\Box h_{\kappa\lambda}\,
\Big)\qquad \mbox{where}\qquad h=\eta^{\mu\nu}h_{\mu\nu}\qquad\mbox{and}\qquad\Box=\partial_\mu\partial^\mu
\end{equation}
Consequently the Einstein tensor is
\begin{equation}\label{og06}
\mathcal G_{\mu\nu}\,=\,\frac{1}{2}\,\tau^{\kappa\lambda}\,\left[
\,2
\,\partial_\kappa\partial_{(\mu} h_{\nu)\lambda}-\partial_\mu\partial_\nu h_{\kappa\lambda}-\partial_\kappa\partial_\lambda h_{\mu\nu}
-\frac{1}{2}\eta_{\mu\nu}\,\Big(\,
2\,\partial_\lambda\partial_\alpha h^\alpha_{\;\;\kappa}-\partial_\kappa\partial_\lambda h-\Box h_{\kappa\lambda}\,
\Big)
\right].
\end{equation}
In order  to simplify the Einstein tensor, we impose several gauge constraints to eliminate non physical degrees of freedom. In order to eliminate the new solutions that are simple coordinate changes, we consider the general coordinate transformation
\begin{equation}\label{og07}
x^\mu\;\to\; x'^\mu=x^\mu+\xi^\mu,\qquad\mbox{where}\qquad\left|\xi^\mu(\bm x)\right|\ll 1.
\end{equation}
The metric in the new coordinate system is obtained from
\begin{equation}\label{og08}
g'^{\mu\nu}=g^{\kappa\lambda}\,\frac{\partial x'^\mu}{\partial x^\kappa}\,\frac{\partial x'^\nu}{\partial x^\lambda}
\end{equation}
so that
\begin{equation}\label{og09}
g'_{\mu\nu}\,=\,\eta_{\mu\nu}+h'_{\mu\nu}\qquad\mbox{where}\qquad h'_{\mu\nu}=h_{\mu\nu}-\partial_\mu\xi_\nu-\partial_\nu\xi_\mu.
\end{equation}
Consequently, 
\begin{equation}\label{og10}
\Gamma'^\lambda_{\mu\nu}\,=\,\Gamma_{\mu\nu}^\lambda\,-\,\tau^{\lambda\kappa}\partial_\mu\partial_\nu\xi_\kappa
\end{equation}
and the equations of motion are invariant to this transformation, as it is expected to be. In order to obtain a specific gauge condition to simplify the Einstein tensor and find the field equations, we impose the constraint
\begin{equation}\label{og0100}
\tau^{\mu\nu}\Gamma^\lambda_{\mu\nu}\,=\,0,
\end{equation}\label{og010}
that is similar to the harmonic coordinate system \cite{Weinberg:1972kfs}. Consequently, the gauge condition will be
\begin{equation}\label{og11}
2\,\tau^{\kappa\lambda}\,\partial_\kappa\partial_{(\mu} h_{\nu)\lambda}-\partial_\mu\partial_\nu \tau\cdot h=0
\qquad\mbox{where}\qquad \tau\bm\cdot h=\tau^{\kappa\lambda}h_{\kappa\lambda}.
\end{equation}
Using (\ref{og11}) and defining 
\begin{equation}\label{og12}
\tau\cdot\partial^2\;=\;\tau^{\mu\nu}\partial_\mu\partial_\nu\;=\;\frac{1}{c^2}\,\partial^2_t+\nabla^2
\end{equation}
we obtain the modified Einstein tensor
\begin{equation}\label{og13}
\mathcal G_{\mu\nu}\,=\,-\frac{1}{2}\,\tau\cdot\partial^2\left[\,h_{\mu\nu}
-\frac{1}{2}\eta_{\mu\nu} h\right].
\end{equation}
We point out that $\,\tau\cdot\partial^2\,$ is identical to the Laplace operator in a four dimensional plane Riemannian metric. We avoid this terminology here because the metric is of course pseudo-Riemannian and the definition of the Laplacian is also different of it. In the usual solution of GW, the $\,\tau\cdot\partial^2\,$ operator is of course the D'Alembertian operator $\big[\,\Box\;\; \mbox{in}\;\; (\ref{og05})\big]$, and the single difference between them is a signal flip on the time second derivative.
The gauge constraint and the Einstein tensor must accordingly be gauge invariant. Thus, using (\ref{og09}) and (\ref{og11})
we get 
\begin{equation}\label{og14}
\tau\cdot\partial^2 \partial_{(\mu}\xi_{\nu)}=0,\qquad\mbox{and thus}\qquad \tau\cdot\partial^2\xi_\mu=0.
\end{equation}
In the Hilbert gauge, the Einstein equations is thus
\begin{equation}\label{og150}
-\frac{1}{2}\,\tau\cdot\partial^2 p_{\mu\nu}=\kappa T_{\mu\nu}\qquad\mbox{for}\qquad p_{\mu\nu}\,=\,h_{\mu\nu}
-\frac{1}{2}\eta_{\mu\nu} h.
\end{equation}
We point out that we can also use (\ref{og02}) and obtain
\begin{equation}\label{og140}
-\frac{1}{2}\,\tau\cdot\partial^2 h_{\mu\nu}=\kappa S_{\mu\nu}.
\end{equation}
In the empty space, where $\,T_{\mu\nu}=0,\,$ we have
\begin{equation}\label{og15}
\tau\cdot\partial^2 h_{\mu\nu}=0.
\end{equation}
Hence the linearized equation set for the transformed gravitational waves is complete and the analogy to the usual linear approximation is exact, the differences being the gauge choice and the operator that acts over the perturbation field $\,h_{\mu\nu}.$ In the next section we seek the simplest possible solutions of this model.

\section{\;\sc Empty space solution\label{PS}}
In the usual  non transformed case, the field equations to the empty space reads
\begin{equation}\label{ps1000}
\Box h_{\mu\nu}=0,
\end{equation}
and is solved in terms complex exponential functions \cite{Weinberg:1972kfs}
\begin{equation}\label{ps1001}
h_{\mu\nu}\big(\bm x,\, t\big)\,=\,\varepsilon_{\mu\nu}\,e^{ik_\mu x^\mu}\,+\,\varepsilon^*_{\mu\nu}\,e^{-ik_\mu x^\mu}
\end{equation}
Where $\,k_\mu\,$ is the momentum four-vector, $\,\varepsilon_{\mu\nu}\,$ is the polarization tensor, and $\,\varepsilon^*_{\mu\nu}\,$ is its complex conjugate. This is the so called plane wave solution. In the transformed case, let us entertain a two dimensional case, where the time $\,t\,$ and the space $\,z\,$ coordinates are changed to
\begin{equation}\label{ps000}
u=c t + z,\qquad\qquad v=c t - z,
\end{equation}
and the two-dimensional field equation (\ref{og15}) accordingly reads
\begin{equation}\label{ps1002}
\frac{\partial^2 h_{\mu\nu}}{\partial u^2}\,+\,\frac{\partial^2 h_{\mu\nu}}{\partial v^2}\,=\,0.
\end{equation}
A separate variables solution to (\ref{ps1002}) is 
\begin{equation}\label{ps00}
h_{\mu\nu}\big(u,\,v\big)\,=\,\left(\varepsilon_{\mu\nu}\,e^{ik v}\,+\,\varepsilon^*_{\mu\nu}\,e^{-ik v}\right)\,\left(A\,e^{ku}\,+\,B\,e^{-ku}\right),
\end{equation}
where $A$ and $B$ are  real integration constants. The usual plane wave solution (\ref{ps1001}) involves complex exponential functions, and the novelty of (\ref{ps00}) are the real exponential functions, whose diverging or evanescent behavior suggests that the physical solutions  involve $A=0$ to have finite solutions and keep $\,|h_{\mu\nu}|\ll 1.\,$ However,  $u\to 0$ when $\,t,z\to\infty\;$ if $\;z>0,\,$ and thus non-evanescent finite solutions are possible in specific situations as well. Irrespective of its asymptotic behavior, the simplest two-dimensional solution of (\ref{og15}) is
\begin{equation}\label{ps001}
h_{\mu\nu}\big(z,\,t\big)\,=\,\left(\varepsilon_{\mu\nu}\,e^{ik z}\,+\,\varepsilon^*_{\mu\nu}\,e^{-ik z}\right)\,e^{-ckt},
\end{equation}
where $\,k\,$ is identified to the momentum transported by the wave,  and because of the exponential decaying we can call such transformed solution as attenuated solution. The above function also satisfies the heat-like equations
\begin{equation}\label{ps01}
\frac{\partial h_{\mu\nu}}{\partial t}\,=\,\frac{c}{k}\nabla^2 h_{\mu\nu}
\end{equation}
and indicates that the propagation of this sort of gravitational wave takes benefit from small momenta. The smaller the momentum, the greater the dispersion of the wave.  From a physical point of view, the gravitational propagation obeys a heat-like propagation pattern, where the space-time can be compared to a material medium of diffusivity constant $\,c/k.\,$
On the other hand, (\ref{ps001}) can also be obtained from the usual  solution (\ref{ps1001}) by way of the symmetry transformation
\begin{equation}\label{ps0010}
t\to -it \qquad\Rightarrow\qquad e^{-ik_\mu x^\mu}\to e^{i kz-ckt},
\end{equation}
that is identical to the Wick (or Euclidean) rotation, commonly employed for path integrals \cite{Das:2019jmz}. This transformation  will be further discussed in the description of the gravitational wave generated from a source in the next section. The solution (\ref{ps001})  also satisfies the field equation after exchanging the time and spatial coordinates, and the oscillation and the attenuation have different physical interpretations in this situation. However, this discussion must be held considering specific examples, something that we leave as a direction for future research. Irrespective of the physical context, using the solution (\ref{ps001}) and the gauge constraint (\ref{og11}), we get
\begin{equation}\label{ps02}
\varepsilon_{0\mu}\,=\,\varepsilon_{3\mu}\,=\,0.
\end{equation}
Consequently, three degrees of freedom stay associated to $\varepsilon_{11},\,\varepsilon_{12},\,$ and $\,\varepsilon_{22}.\,$
The transformed gauge condition is tighter than the customary gauge condition of the non transformed case, that keeps six polarization directions from the originally ten degrees of freedom of the unconstrained polarization tensor. A remaining non-physical polarization direction is eliminated using the coordinate gauge (\ref{og0100}), so that
\begin{equation}\label{ps03}
 \varepsilon_{11}\,=\,-\,\varepsilon_{22}.
\end{equation}
In agreement to  the non-transformed case, the physically relevant polarizations are two, and these two remaining polarization
directions are those that have helicity number equal two \cite{Weinberg:1972kfs}. This result open the possibility that the  attenuated and non-attenuated GW could interact, and this is an interesting idea for future research.

\subsection{\; Energy momentum tensor}

We can obtain the the transformed energy-momentum tensor using the usual GW method. The transformed Ricci tensor expanded in terms of $\,h_{\mu\nu}\,$ is
\begin{equation}\label{ps04}
\mathcal R_{\mu\nu}\,=\,\mathcal R_{\mu\nu}^{(1)}\;+\,\mathcal R_{\mu\nu}^{(2)}\,+\,\mathcal O\left(h^3\right)
\end{equation}
where
\begin{eqnarray}\label{ps05}
\mathcal R_{\mu\kappa}^{(1)}&=&\frac{1}{2}\eta^{\lambda\nu}\tau_\lambda^{\;\;\;\lambda'}
\Bigg(\partial^2_{\kappa\mu}h_{\lambda'\nu}\,-\,\partial^2_{\kappa\lambda'}h_{\mu\nu}\,-\,\partial^2_{\mu\nu}h_{\kappa\lambda'}\,+\,\partial^2_{\lambda'\nu}h_{\kappa\mu}\,\Bigg)\\
\nonumber
\mathcal R_{\mu\kappa}^{(2)}&=&\frac{1}{2} h^{\lambda\nu}\tau_\lambda^{\;\;\;\lambda'}
\Bigg(\partial^2_{\kappa\mu}h_{\lambda'\nu}\,-\,\partial^2_{\kappa\lambda'}h_{\mu\nu}\,-\,\partial^2_{\mu\nu}h_{\kappa\lambda'}\,+\,\partial^2_{\lambda'\nu}h_{\kappa\mu}\,\Bigg)\,
+\,\frac{1}{4}\eta^{\lambda\nu}\eta^{\rho\sigma}\Bigg(\Gamma_{\kappa\rho\lambda}\Gamma_{\mu\sigma\nu}\,-\,
\Gamma_{\nu\rho\lambda}\Gamma_{\mu\sigma\kappa}\Bigg)
\end{eqnarray}
and $\,\Gamma_{\nu\rho\lambda}\,=\,\eta_{\nu\nu'}\Gamma^{\nu'}_{\rho\lambda}.\,$ From (\ref{i00}) and (\ref{ps04}-\ref{ps05}), we get Ricci scalar
\begin{equation}\label{p06}
\mathcal R\,=\,\mathcal R^{(1)}\,+\,\mathcal R^{(2)}\,+\,h^{\mu\kappa}\mathcal R_{\mu\kappa}^{(1)}\qquad\mbox{with}\qquad
\mathcal R^{(i)}\,=\,\eta^{\mu\kappa}\mathcal R_{\mu\kappa}^{(i)}.
\end{equation}
Keeping the first order terms of the Riemann expansion in the Einstein tensor and absorbing the higher order terms into the energy-momentum tensor, the modified Einstein equations become
\begin{equation}\label{ps07}
\mathcal G_{\mu\kappa}^{(1)}=\kappa\left(T_{\mu\kappa}+t_{\mu\kappa}\right)
\end{equation}
where
\begin{eqnarray}\label{ps08}
\mathcal G_{\mu\kappa}^{(1)}&=&\mathcal R_{\mu\kappa}-\frac{1}{2}\eta_{\mu\kappa}\mathcal R^{(1)}\\
t_{\mu\kappa}&=&\frac{1}{2\kappa}\Bigg[\,h_{\mu\kappa}\mathcal R^{(1)}\,+\,\eta_{\mu\kappa}\left(h^{\rho\sigma}\mathcal R_{\rho\sigma}^{(1)}\,+\,\mathcal R^{(2)}\right)\,-\,2\,\mathcal R_{\mu\kappa}^{(2)}\,\Bigg].
\end{eqnarray}
This analysis is of course usual for the linearized approximation of GR \cite{Weinberg:1972kfs}, the only difference is the appearance of the $\,\tau_{\mu\nu}\,$ tensor in specific places. In analogy to the usual case, the gauge condition (\ref{og11}) and the equation of motion (\ref{og15}) imply that
\begin{equation}\label{ps09}
\mathcal R_{\mu\nu}^{(1)}\,=\,0.
\end{equation}
The effective energy momentum tensor on the right hand side of (\ref{ps07}) is also zero, and thus
\begin{equation}\label{ps10}
T_{\mu\kappa}=\frac{1}{\kappa}\left(\mathcal R^{(2)}_{\mu\kappa}\,-\,\frac{1}{2}\eta_{\mu\kappa}\mathcal R^{(2)}\,\,\right).
\end{equation}
An explicit form of (\ref{ps10}) is achieved from the polarization condition (\ref{ps02}) and the independence from coordinates $x$ and$y$, so that
\begin{eqnarray}\nonumber
\mathcal R^{(2)}_{\kappa\mu}&=&\,\frac{1}{4}\eta^{\lambda\nu}\eta^{\rho\sigma}\Bigg(\,
\partial_\kappa h_{\lambda\rho}\partial_\mu h_{\nu\sigma}\,+\,2\,\partial_\lambda h_{\kappa\rho}\partial_\nu h_{\mu\sigma}\,-\,
\partial_\rho h_{\nu\lambda}\partial_\sigma h_{\kappa\mu}\,\Bigg)\\
\label{ps11}
\mathcal R^{(2)}&=&\,\frac{3}{4}\,\partial_\lambda h_ {\rho\sigma}\partial^\lambda h^{\rho\sigma}\,-\,\frac{1}{4}\,\partial_\lambda h\,\partial^\lambda h,
\end{eqnarray}
and $\,h=\eta^{\rho\sigma}h_{\rho\sigma}.\,$ Finally, (\ref{ps001}) and ({\ref{ps03}) imply that $\,h=0\,$ and therefore
\begin{eqnarray}\nonumber
\mathcal R^{(2)}_{00}&=&k^2\left(\,\big|\varepsilon_{12}\big|^2+\big|\varepsilon_{11}\big|^2+
\frac{\varepsilon_{11}^2+\varepsilon_{12}^2}{2}\,e^{2ikz}\,+\,
\frac{\varepsilon^{*\;2}_{11}+\varepsilon^{*\;2}_{12}}{2}\,e^{-2ikz}
\right)e^{-2ckt}\\
\nonumber
\mathcal R^{(2)}_{33}&=&k^2\left(\,\big|\varepsilon_{12}\big|^2+\big|\varepsilon_{11}\big|^2-
\frac{\varepsilon_{11}^2+\varepsilon_{12}^2}{2}\,e^{2ikz}\,-\,
\frac{\varepsilon^{*\;2}_{11}+\varepsilon^{*\;2}_{12}}{2}\,e^{-2ikz}
\right)e^{-2ckt}\\
\label{ps12}
\mathcal R^{(2)}_{03}&=&k^2\left(
\frac{\varepsilon_{11}^2+\varepsilon_{12}^2}{2i}\,e^{2ikz}\,-\,
\frac{\varepsilon^{*\;2}_{11}+\varepsilon^{*\;2}_{12}}{2i}\,e^{-2ikz}
\right)e^{-2ckt}\\
\nonumber
\mathcal R^{(2)}_{11}&=&-\,k^2\left(
\frac{\varepsilon_{11}^2+\varepsilon_{12}^2}{2}\,e^{2ikz}\,+\,
\frac{\varepsilon^{*\;2}_{11}+\varepsilon^{*\;2}_{12}}{2}\,e^{-2ikz}
\right)e^{-2ckt}\\
\nonumber
\mathcal R^{(2)}_{22}&=&\mathcal R^{(2)}_{11},\qquad\qquad\mathcal R^{(2)}_{12}\,=\,0,\qquad\qquad \mathcal R^{(2)}\,=\,3\,\mathcal R^{(2)}_{11}.
\end{eqnarray}
Discarding the complex exponential terms by averaging over a period oscillation, the expectation values of the components of the energy momentum tensor comply with
\begin{equation}\label{ps13}
\big\langle T_{00}\big\rangle\,=\,\frac{k^2}{\kappa}\left(\,\big|\varepsilon_{12}\big|^2+\big|\varepsilon_{11}\big|^2\right)e^{-2ckt},\qquad
\big\langle T_{00}\big\rangle\,=\,\big\langle T_{33}\big\rangle\qquad\mbox{and}\qquad\big\langle T_{\mu\nu}\big\rangle\,=\,0\qquad\mbox{otherwise}.
\end{equation}
The agreement to the usual case is perfect, and the unique difference is the real decaying exponential on the right hand side of (\ref{ps13}).
We point out that exponential decaying is not a novelty in gravitation, and, by way of example, the Yukawa potential \cite{Ohanian:1995uu} found several applications in gravity \cite{Berezhiani:2009kv,Mazharimousavi:2019ksw}. The momentum  $k$ transported by the gravitational wave plays a role analogue to the range $\lambda$ of the Yukawa potential, and the possible relation between both of the attenuation parameters would be interesting direction for future research. Finally, attenuated gravitational  radiation is also not a novelty, because of the proposed cosmological damping \cite{Weinberg:2004dgr}, and several interesting ways of ascertaining this sort of effect are summarized in \cite{Saltas:2014dha}, although the attenuated gravitational radiation with geometrical origin introduced in this article is certainly  original.
\section{\;\sc Emission of gravitational waves\label{E}}
In the presence of a gravitational source, the energy-momentum tensor is non trivial and the field equations (\ref{og140}) read
\begin{equation}\label{e01}
\frac{1}{c^2}\,\partial^2_t\,h_{\mu\nu}+\nabla^2 h_{\mu\nu}\,=\,-\,2\kappa S_{\mu\nu}.
\end{equation}
Solutions of (\ref{e01}) can be generated from the solutions of $\Box h_{\mu\nu}=2\kappa S_{\mu\nu}$ exploiting the transformation (\ref{ps0010}) and the energy-momentum tensor written in terms of a Fourier transform
\begin{equation}\label{e02}
T_{\mu\nu}(\bm x, \,t)\,=\,\int_0^\infty T_{\mu\nu}(\bm x, \,\omega) e^{-i\omega t} d\omega \;+\; \mbox{c. c.}	
\end{equation}
Restraining the solution to the wave zone, the distances from the source are much larger than the dimension $\,|\bm x'|_{max}\,$ of the source. If $\,\bm x'\,$ is the coordinate of the inner points of the source, in the wave zone holds 
\begin{equation}\label{e03}
|\bm x|\,\gg\, |\bm x'|_{max}\qquad\mbox{as well as}\qquad|\bm x|\,\gg\, \omega |\bm x'|^2_{max}\qquad\mbox{and}\qquad|\bm x|\,\gg\, \frac{1}{\omega}.
\end{equation}
Thus, the solutions obey the same structure as (\ref{ps001}), but the space dependence changes as $\,z\to\bm x\,$ and the polarization tensor is
\begin{equation}
\varepsilon_{\mu\nu}\,=\,\frac{\kappa c^4}{2\pi}\frac{1}{|\bm x|}\int T_{\mu\nu}(\bm x',\,\omega) e^{-i\bm{k\cdot x}'}d^3x'\qquad
\mbox{where}\qquad \bm k=\omega\frac{\bm x}{|\bm x|}.
\end{equation}
Following the ordinary non-transformed case \cite{Weinberg:1972kfs}, the emitted power per solid angle in a direction $\bm x/|\bm x|$ is
\begin{equation}
\frac{dP}{d\Omega}\,=\,-\,|\bm x|^2 x_i\left\langle t^{i0}\right\rangle
\end{equation}
and the sole difference from the usual case is the attenuation due to the real exponential factor, as is seen in (\ref{ps13}). 

\section{\sc Conclusion}

In this paper we obtained a way for generating gravitational waves that combined the local perturbation used in the standard GW description (\ref{i00}) with a global transformation of the affine connection (\ref{i03}). In other words, an algebraic transformation was introduced into the geometric procedure. Although the method was previously deployed within the gravito-electromagnetic approximation \cite{Giardino:2018ffd,Giardino:2021gwq}, the algebraic transformation was not explicitly described in that article.  

This new gravitational radiation contains an attenuation factor in the form of a real exponential function, something that is not observed in the usual linear approximation description of GW. Additionally, the gauge constraint is tighter, and the polarization directions are only two.
One may inquire which of the gravitational radiations are actually generated, but it seems reasonable to expect that both of the processes can be ascribed to a physical source of oscillations in the space-time, and even that interaction between both of the processes must happen. We also stress that the attenuation of gravity has already been introduced in static gravity in the work of Yukawa potentials  \cite{Goldhaber:2008xy}, and these effects can be observed, for example, if gravitational waves were naively detected separated different points of the space. Sophisticated and more realistic ways of studying gravitational attenuated phenomena have been proposed in \cite{Saltas:2014dha}, while numerical work can be used to determine the suitability of the model in describing gravitational wave phenomena generated by compact-object binaries \cite{Toffano:2019ekp}.  Despite these possibilities, the experimental determination of evanescent gravitational waves is a fundamental issue in order that the theory presented in this article be considered seriously. While we do not have it, we have to acknowledge this experimental blank as a rationale to discard gravitational theories such as those generated by the transformations (\ref{i0005}-\ref{Rie}).

The future directions are various, not only the application to the well known results of GW \cite{Maggiore:1900zz,Nielsen:2019caa}, but also modeling the novel possibilities that are not found in the usual theory, like an attenuated wave amplitude.  A further important investigation concerns the investigation of the general covariance, whether the transformation (\ref{i02}) does not implies in the breakdown of the Lorentz invariance. We understand that this study can be entertained using the language of fiber bundles, vierbeine and spin connection.  A phenomenological study can be determining the upper limit of the dissipation of GW \cite{Loeb:2020apj},  and another interesting direction concerns inquiring whether (\ref{i01}) can be generalized to more general algebraic  transformations that could generate oscillations of the space-time.
A further interesting possibility is to study quantum effects using the attenuation factor, something that has already been done in the  case of the Yukawa theory \cite{Goldhaber:2008xy}. Finally, the relation of the presented results and the affine gravity theories \cite{Hehl:1994ue} generated by a possible nonmetricity character (\ref{i0005}) in (\ref{AC}-\ref{Rie}) is also another exciting direction for future research.

\section*{Declarations}

\paragraph{Acknowledgments} The author gratefully thanks the anonymous reviewer by pointing out the discussion of the nonmetricity constraint that appeared in (\ref{i0005}-\ref{Rie}), and the relation of the proposal of the paper to the metric-affine gravity.

\paragraph{Data availability statement}The author declares that data sharing is not applicable to this article as no datasets were generated or analysed during the current study.

\paragraph{Conflict of interests statement} There are no known competing financial interests or personal relationships that
could have appeared to influence the work reported in this paper.

%
%
%
%

\begin{footnotesize}
\bibliographystyle{unsrt} 

\end{footnotesize}

\end{document}